# Magnetoelectric near fields

Eugene Kamenetskii

Microwave Magnetic Laboratoty
Ben Gurion University of the Negev, Israel
E-mail: kmntsk@bgu.ac.il

**Abstract**

Similar to electromagnetic (EM) phenomena, described by Maxwell equations, physics of magnetoelectric (ME) phenomena deals with the fundamental problems of the relationship between electric and magnetic fields. The different nature of these two notions is especially evident in dynamic regimes. Analyzing the EM phenomena inside the ME material, the question arises: What kind of the near fields, originated from a sample of such a material, can be measured? Observation of the ME states requires an experimental technique characterized by a violation of spatial and temporal inversion symmetries in a subwavelength region. This presumes the existence of specific near fields. Recently, such field structures, called ME fields, were found as the near fields of a quasi-2D subwavelength-size ferrite disk with magnetic-dipolar-mode (MDM) oscillations. The key physical characteristics that determine the configurations of the ME near fields are the spin and orbital angular momenta of the quantum states of the MDM spectra. This leads to the appearance of subwavelength power-flow vortices. By virtue of unique topology, the ME quantum fluctuations in vacuum are different from virtual EM photons. While preserving the ME properties, one observes strong enhancing the near-field intensity. The main purpose of this chapter is to review and analyze the studies of the ME fields. We consider the near-field topological singularities originated from the MDM ferrite-disk particle. These topological features can be transmitted to various types of nonmagnetic material structures.

## 1. Introduction

Symmetry principles play an important role in the laws of nature. Maxwell added an electric displacement current to put into a symmetrical form the equations which couple together the electric and magnetic fields. The dual symmetry between electric and magnetic fields underlies the conservation of energy and momentum for electromagnetic fields [1]. Recently, it was shown that this dual symmetry determines the conservation of optical (electromagnetic) chirality [2, 3]. Based on an analysis of the interaction of chiral light and chiral specimens, new mechanisms of enantiomer discrimination and separation in optics have been proposed [2 – 8]. Since chiroptical effects are usually hampered by weak chiral light-matter interaction, it is argued that to enhance the chiral effects it is necessary that the near field remains chiral in the process. Different plasmonic and dielectric nanostructures have recently been proposed as a viable route for near-field enhancement of chiral light-matter interactions [9 – 11]. In a more general sense, one can say that this is an attempt to enhance the near-field intensity while preserving the ME properties. However, the following fundamental questions arise: Can one really observe effects of the near-field magnetoelectricity in dynamic regimes. What are the symmetry properties of such dynamic ME fields?

    The question on relationships between magnetoelectricity and electromagnetism is a subject of a strong interest and numerous discussions in microwave and optical wave physics and material

sciences. The problem of the near-field magnetoelectricity in electromagnetism is a topical problem. Evanescent fields are oscillating fields whose energy is spatially concentrated in the vicinity of the oscillating currents. In classical electrodynamics we know only two types of local (subwavelength) electric currents: linear and circular. These types of currents determine elementary electric and magnetic dipole oscillations in matter. The electric polarization is parity odd and time-reversal even. At the same time, the magnetization is parity even and time-reversal odd [1]. These symmetry relation cast doubt on the idea of a local (subwavelength) coupling of two, electric and magnetic, small dipoles. When the violation of the invariances under space reflection parity and time inversion are necessary conditions for the emergence of the ME effect, the same symmetry properties should be observed for the near fields – the ME near fields.

The uniqueness of the proposed ME near fields can be shown by analyzing vacuum near fields originated from a scatterer made of a ME structure. In this connection, it is worth noting that in a case of usual (non-ME) material structures one can distinguish two kinds of the EM near fields: (a) near fields originated from EM wave resonances and (b) near fields originated from dipole-carrying resonances. The former fields, abbreviated as EM NFs, are obtained based on the full-Maxwell-equation solutions with use of Mie theory [12]. The latter fields, abbreviated as DC NFs, are observed when the electric or magnetic dipole-carrying oscillations (such, for example, as surface plasmons [13 – 15] and magnons [16 – 18]) take place. Notably, in accordance with Mie theory one can observe EM NFs with magnetic responses originated from small nonmagnetic dielectric resonators, both in microwaves [19] and optics [20 – 22]. In a case of DC NFs, strong coupling of EM waves with electric or magnetic dipole-carrying excitations, called polaritons, occur [16, 23]. Importantly, the spatial scale of the DC NFs is much smaller than the spatial scale of the EM NFs in the same frequency range. Due to strong coupling of EM waves with dipole-carrying excitations and temporal dispersion of the material, polaritons display enhanced field localization to surfaces and edges. Properties of vacuum near-fields originated from a small non-ME (dielectric or magnetic) sample become evident when this sample has sizes significantly smaller than the EM wavelength *in all three spatial dimensions*. The matter of fact is that near such a scatter we can only measure the electric $\vec{E}$ or the magnetic field $\vec{H}$ with accuracy. As volumes smaller than the wavelength are probed, measurements of EM energy become uncertain, highlighting the difficulty with performing measurements in this regime. There is Heisenberg's uncertainty principle binding $\vec{E}$ and $\vec{H}$ fields of the EM wave [13, 24].

Taking all this into account, let us consider now a *subwavelength ME sample*. The near-field structure of such a point scatterer is dominated by two types of the fields: the electric and magnetic fields, which are *mutually coupled* due to the intrinsic properties of a ME material. This fact gives us much greater uncertainty in probing of the fields. In total, such fields can be represented as the structures of cross $\vec{E} \times \vec{H}$ or dot $\vec{E} \cdot \vec{H}$ products in a subwavelength region. Due to $\mathcal{PT}$ symmetry of ME structure, the ME near fields of a subwavelength sample should be characterized by a certain pseudoscar parameter. Moreover, supposing that in a subwavelength region *both* structures of cross $\vec{E} \times \vec{H}$ and dot $\vec{E} \cdot \vec{H}$ products exist, one should assume the presence of helicity properties of the fields. It is evident that such a near-field structure – the ME-field structure – is beyond the frames of the Maxwell theory description [1].

When we are talking on ME dynamics, we have to refer also to an analysis of artificial structures – bianisotropic metamaterials. The notion "bianisotropic media" had been introduced to generalize different effects of coupling between magnetic and electric properties [25]. The local bianisotropic media is supposed as the media composed by structural subwavelength elements with "glued" pairs of electric and magnetic dipoles. The consideration of high-order quadrupole and multipole transitions is actually an account of spatial dispersion [26, 27]. It is assumed that bianisotropy (chirality) in metamaterials arises from a "local ME effect" [28 – 31]. Such a "first-principle", "microscopic-scale" ME effect of a structure composed by "glued" pairs of electric and magnetic dipoles raises a basic question on the ways of probing the dynamic parameters, since the near field structure of such a probe should violate both the spatial and temporal inversion symmetries. However, in metamaterial bianisotropic (chiral) structures, the known experimental retrieval of the cross-polarization parameters is via *far-field measurement* of the scattering-matrix characteristics [32 – 34]. Far-field retrieved permittivity and permeability frequently retain non-physical values, especially in the regions of the metamaterial resonances where most interesting features are expected. Far-field retrieved cross-polarization parameters of "bianisotropic particles" retain much greater non-physical value. The observed far-field phenomena of bianisotropy (chirality) can be very weakly related to the near-field manipulation effects. We can say that the cross-polarization properties of small "glued-pair" bianisotropic particles are incompatible with the effects of Rayleigh scattering.

In this chapter, we consider near fields originated from subwavelength resonators, that are the systems with quantum-confinement effects of dipolar-mode quasistatic oscillations. We analyze the possibilities of these resonances to exhibit near-field ME properties. An analysis of such dipole-carrying excitations allows finding a proper way in realizing polariton structures with properties of strong ME interactions.

## 2. Subwavelength resonators with dipole-carrying excitations

An interaction between the photon and medium dipole-carrying excitation becomes strong enough near the resonance between the light mode and the mode of the medium excitation. At the resonance region, the dispersion curves of these modes transform into two split polaritonic branches showing anticrossing behavior. The examples are exciton polaritons, surface-plasmon polaritons, and magnon polaritons. Semiclassically, polaritons are described using Maxwell equations and constitutive relations that include the frequency dependent response functions. Quantum mechanically, polaritons are described as hybrid collective excitations that are linear superpositions of matter collective excitations and photons. There are the effects of *interaction between real and virtual photons*. When dipole-carrying excitations are observed in a high-quality confined structure, the coupling modes can appear as composite bosons. Strong long-range dipole-dipole interactions significantly modify the mean-field predictions of the quantum phases of microscopic short-range excitations by stabilizing the condensate phase. It can persist up to densities high enough to support quantum liquidity with very long lifetimes. In *exciton-polariton condensates*, in particular, this effect leads to sustained trapping of the emitted photon [35 – 39].

Excitons in semiconductor resonators are dipole-carrying oscillations. Plasmons and magnons in confined structures are also dipole-carrying oscillations. Plasmons are optical responses of metal structures arising from collective oscillations of their conduction electrons. The microwave responses of ferrite samples – magnons – arise from collective oscillations of their precessing electrons. Both plasmons and magnons are bosons. In increasing the capabilities of the optical and microwave techniques further into the subwavelength regime, small plasmon and magnon resonant structures have attracted considerable interest. These oscillations in subwavelength resonators, however, are not *composite bosons*, as in the case of exciton resonances. No dipole-dipole plasmon condensate and dipole-dipole magnon condensate in confined resonant structures are observed, to the best to our knowledge. The problem of creating a condensate with linked (electric *and* magnetic) dipole-carrying excitations confined in a high-quality resonant structure appears as a lot more exotic. Is it even possible to observe tightly bound ME excitations, which turns into a composite boson (or fermion) and behaves as quasiparticle? Can we, in general, solve the problem of creation of the ME-polariton condensate? To answer these questions, we should analyze the possibility of finding ME properties in subwavelength resonators with quasistatic (dipolar-mode) oscillations.

### A. On the possibility to observe the quantum confinement effects of electrostatic and magnetostatic oscillations

Electromagnetic (EM) responses of plasmon oscillations in optics and magnon oscillations in microwaves give rise to a strong enhancement of local fields near the surfaces of subwavelength resonators. We can classify these oscillations as electrostatic (ES) and magnetostatic (MS) resonances, respectively [18, 40]. In ES resonances in small metallic samples, one neglects a time variation of magnetic energy in comparison with a time variation of electric energy. It means that one neglects a magnetic displacement current and an electric field is expressed via an ES potential, $\vec{E} = -\vec{\nabla}\phi$ [40]. However, the Ampere–Maxwell law gives the presence of a curl magnetic field. In like manner, in the case of MS resonances in ferrite samples, one neglects a time variation of electric energy in comparison with a time variation of magnetic energy. It means that the MS-resonance problem is considered as zero-order approximation of Maxwell's equations when one neglects the electric displacement current and expresses a magnetic field via a MS potential, $\vec{H} = -\vec{\nabla}\psi$ [18]. While Faraday's law gives the presence of a curl electric field. Importantly, from a classical electrodynamics point of view, one does not have a physical mechanism describing the reverse effect of transformation of a curl magnetic field to a potential electric field in the case of ES resonances. Also, one does not have a physical mechanism describing the reverse effect of transformation of a curl electric field to a potential magnetic field a case of the MS resonance [1, 41]. It means that, fundamentally, subwavelength sizes of the particles should eliminate any EM retardation effects. We can say that for an EM wavelength $\lambda$ and particle of a characteristic size $a$, the quasistatic approximation $2\pi a/\lambda \ll 1$ implies the transition to a small EM phase.

What kind of the time-varying field structure one can expect to see when an electric or magnetic displacements currents are neglected and so the electromagnetic-field symmetry (dual symmetry) of Maxwell equations is broken? When one neglects a displacement current (magnetic or electric) and considers the scalar-function [$\phi(\vec{r},t)$ or $\psi(\vec{r},t)$] solutions, as the wave-propagation solutions, one has to accept the possibility to observe the quantum confinement effects of electrostatic and magnetostatic oscillations. Such an analysis of quasistatic resonances is based on postulates about a physical meaning of scalar function as a complex scalar wavefunction, which presumes a long-range phase coherence in dipole–dipole interactions. These solutions should be based on the Schrödinger-like equation.

For quasi-ES resonances in subwavelength metal structures characterised by non-homogeneous scalar permittivity, we have Poison's equation [42 – 45]

$$\nabla^2\phi + \vec{\nabla}\cdot\left(\left(\varepsilon(\vec{r})-1\right)\vec{\nabla}\phi\right) = 0. \tag{1}$$

At the same time, for quasi-MS resonances in subwavelength microwave ferrite structures with tensor permeability $\ddot{\mu}$, there is Walker's equation [18, 46]:

$$\nabla^2 \psi + \vec{\nabla} \cdot \left( \left( \frac{\ddot{\mu}}{\mu_0} - \vec{I} \right) \cdot \vec{\nabla} \psi \right) = 0 \,. \tag{2}$$

Solutions of both these equations are harmonic functions. Nevertheless, it appears that in spite of a certain similarity between Eqs. (1) and (2), the physical properties of the ES and MS oscillation spectra are fundamentally different in many aspects. The most important factor distinguishing the MS resonance from the ES resonance is the tensorial form of permeability and the presence off-diagonal gyrotropic elements in this tensor.

In Ref. [47] it was discussed that in a case of the surface plasmon resonances in subwavelength optical metallic structures no retardation processes characterized by the electric dipole-dipole interaction and described exclusively by electrostatic wave function $\phi(\vec{r},t)$ take place. There is no possibility to describe these resonances by the Schrödinger-equation energy eigenstate problem. Nevertheless, for MS resonances in ferrite specimens we have bulk wave process, which are determined by a scalar wave function $\psi(\vec{r},t)$. Due to the retardation processes caused by the magnetic dipole-dipole interaction in a subwavelength ferrite particle, we have a possibility to formulate the energy eigenstate boundary problem based on the Schrödinger-like equation for scalar-wave eigenfunctions $\psi(\vec{r},t)$. Such a behavior can be obtained in a ferrite particle in a form of a quasi-2D disk. The oscillations in a quasi-2D ferrite disk, analyzed as spectral solutions for the MS-potential scalar wave function $\psi(\vec{r},t)$, have evident quantum-like attributes. Quantized forms of such oscillations we call the MS magnons or the magnetic-dipolar-mode (MDM) magnons. The macroscopic nature of MDMs, involving the collective motion of a many-body system of precessing electrons, does not destroy a quantum behavior. The long-range dipole-dipole correlation in positions of electron spins can be treated in terms of collective excitations of a system as a whole.

Analyzing the confinement effects of electrostatic and magnetostatic oscillations in subwavelength resonators, it is also worth making another important remark. Considering light interaction with photonic and plasmonic resonances, the authors of review in Ref. [48] noted that as the optical mode becomes deeply subwavelength in all three dimensions, independent of its shape, the Q-factor of the resonances is limited to about 10 or less. As they argue, the reason is that in such small volumes, self-sustaining oscillations are no longer possible between the electric-field and magnetic-field energies and, at the same time, no effects of the electric dipole-dipole oscillations can be assumed. At the same time, in the case of a microwave MDMs in a deeply subwavelength ferrite disk resonator, we have the Q-factor about several thousand [49 – 51]. For such MDM resonances, subwavelength sizes of the ferrite particle allow eliminate any electromagnetic retardation effects and consider only the magnetic dipole-dipole interaction effects.

To make the MDM spectral problem analytically integrable, two approaches were suggested. These approaches, distinguished by differential operators and boundary conditions used for solving the spectral problem, give two types of MDM oscillation spectra in a quasi-2D ferrite disk. These two approaches are conditionally called as the *G* and *L* modes in the magnetic dipolar spectra [52 – 57]. The MS-potential wave function $\psi(\vec{r},t)$ manifests itself in different manners for each of these types of spectra. In the case of the *G*-mode spectrum, where the physically observable quantities are energy eigenstates, the MS-potential wave function appears as a Hilbert-space scalar wave function. In the case of the *L* modes, the MS-potential wave function is considered as a generating function for the vector harmonics of the magnetic and electric fields.

## B. Spectral problems for MDM magnetostatic oscillations: *G* modes

The MDM-resonance spectral solutions obtained from the second-order differential equation – the Walker's equation [18, 46] – are constructed in accordance with basic symmetry considerations for the sample geometry. For an open quasi-2D ferrite disk normally magnetized along the *z* axis, we can use separation of variables. In cylindrical coordinate system $(z, r, \theta)$, the solutions are represented as [52 – 57]

$$\psi_{p,v,q} = A_{p,v,q} \xi_{p,v,q}(z) \tilde{\eta}_{v,q}(r,\theta), \tag{3}$$

where $A_{p,v,q}$ is a dimensional amplitude coefficient, $\xi_{p,v,q}(z)$ is a dimensionless function of the MS-potential distribution along *z* axis, and $\tilde{\eta}_{v,q}(r,\theta)$ is dimensionless membrane function. The membrane function $\tilde{\eta}$ is defined by a Bessel-function order $v$ and a number of zeros of the Bessel function corresponding to a radial variations $q$. The dimensionless "thickness-mode" function $\xi(z)$ is determined by the axial-variation number *p*.

In a quasi-2D ferrite disk, one can formulate the energy eigenstate boundary problem based on the Schrödinger-like equation for scalar-wave eigenfunctions $\psi(\vec{r},t)$ with using the Dirichlet-Neumann (ND) boundary conditions. The energy eigenvalue problem for MDMs is defined by differential equation

$$\hat{G}_{\perp} \tilde{\eta}_n = E_n \tilde{\eta}_n, \tag{4}$$

where $\hat{G}_{\perp}$ is a two-dimensional (on the $r, \theta$ disk plane) differential operator. The quantity $E_n$ is interpreted as density of accumulated magnetic energy of mode *n*. This is the average (on the RF period) energy accumulated in the ferrite-disk region of unit in-plane cross-section and unit length along *z* axis [52 – 57]. The operator $\hat{G}_{\perp}$ and quantity $E_n$ are defined as

$$\frac{g\mu_0}{4} \mu_n \nabla_{\perp}^2 \tilde{\eta}_n = E_n \tilde{\eta}_n, \tag{5}$$

where

$$E_n = \frac{g\mu_0}{4} (\beta_n)^2. \tag{6}$$

Here $\nabla_{\perp}^2$ is the two-dimensional (on the circular cross section of a ferrite-disk region) Laplace operator, $g$ is a dimensional normalization coefficient (with the unit of dimension $\psi^2$) for mode *n* and $\beta_n$ is the propagation constant of mode *n* along the disk axis *z*. The parameter $\mu_n$ (which is a diagonal component of the permeability tensor [18]) should be considered as an eigenvalue. Outside a ferrite $\mu_n = 1$. The operator $\hat{G}_{\perp}$ is a self-adjoint operator only for negative quantities $\mu_n$ in a ferrite.

For self-adjointness of operator $\hat{G}_\perp$, the membrane function $\tilde{\eta}_n(r,\theta)$ must be continuous and differentiable with respect to the normal to lateral surface of a ferrite disk. The homogeneous boundary conditions – the ND boundary conditions – for the membrane function are:

$$\left(\tilde{\eta}_n\right)_{r=\mathcal{R}^-} - \left(\tilde{\eta}_n\right)_{r=\mathcal{R}^+} = 0 \tag{7}$$

and

$$\mu\left(\frac{\partial \tilde{\eta}_n}{\partial r}\right)_{r=\mathcal{R}^-} - \left(\frac{\partial \tilde{\eta}_n}{\partial r}\right)_{r=\mathcal{R}^+} = 0, \tag{8}$$

where $\mathcal{R}$ is the disk radius. MDM oscillations in a ferrite disk are described by real eigenfunctions: $\tilde{\eta}_{-n} = \tilde{\eta}_n^*$. For modes $n$ and $n'$, the orthogonality conditions are expressed as

$$\int_{S_c} \tilde{\eta}_n \tilde{\eta}_{n'}^* dS = \delta_{nn'}, \tag{9}$$

where $S_c$ is a square of a circular cross section of a ferrite-disk region and $\delta_{nn'}$ is the Kronecker delta. The spectral problem gives the energy orthogonality relation for MDMs:

$$\left(E_n - E_{n'}\right) \int_{S_c} \tilde{\eta}_n \tilde{\eta}_{n'}^* dS = 0. \tag{10}$$

Since the space of square integrable functions is a Hilbert space with a well-defined scalar product, we can introduce a basis set. A dimensional amplitude coefficient we write as $A_n = c' a_n$, where $c'$ is a dimensional unit coefficient and $a_n$ is a normalized dimensionless amplitude. The normalized scalar-wave membrane function $\tilde{\eta}$ can be represented as $\tilde{\eta} = \sum_n a_n \tilde{\eta}_n$. The amplitude is defined as $|a_n|^2 = \left|\int_{S_c} \tilde{\eta} \tilde{\eta}_n^* dS\right|^2$. The mode amplitude can be interpreted as the probability to find a system in a certain state $n$. Normalization of membrane function is expressed as $\sum_n |a_n|^2 = 1$ [52 – 57].

The analysis of discrete-energy eigenstates of the MDM oscillations, resulting from structural confinement in a normally magnetized ferrite disk, is based on a continuum model. Using the principle of wave-particle duality, one can describe this oscillating system as a collective motion of quasiparticles. There are "flat-mode" quasiparticles at a reflexively-translational motion behavior between the lower and upper planes of a quasi-2D disk. Such quasiparticles are called "light" magnons. In our study we consider MS magnons in ferromagnet as quanta of collective MS spin waves that involves the precession of many spins on the long-range dipole-dipole interactions. It is different from the short-range magnons for exchange-interaction spin waves with a quadratic character of dispersion.

The meaning of the term "light", used for the condensed MDM magnons, arises from the fact that effective masses of these quasiparticles are much less, than effective masses of "real" magnons – the quasiparticles describing small-scale exchange-interaction effects in magnetic structures. The effective mass of the "light" magnon for a monochromatic MDM is defined as [53]:

$$\left(m_{lm}^{(eff)}\right)_n = \frac{\hbar}{2}\frac{\beta_n^2}{\omega}. \tag{11}$$

In solving boundary value problems for MS resonances, one encounters some questions when using boundary conditions. As is known, in solving a boundary value problem that involves the eigenfunctions of a differential operator, the boundary conditions must be in a definite correlation with the type of this differential operator [58, 59]. In an analysis of MDM resonances in a ferrite disk, we used the homogeneous ND boundary conditions, which mean continuity of the MS wave functions together with continuity of their first derivatives on the sample boundaries. Only in this case the functions form a complete set of orthogonal basis functions and thus the field expansion in terms of orthogonal MS-potential functions can be employed.

However, the ND boundary condition (7), (8) are not the EM boundary conditions. While the considered above ND boundary conditions are the so-called essential boundary conditions, the EM boundary conditions are the natural boundary conditions [58]. For the EM boundary conditions, on a lateral surface of a ferrite disk we have to have continuity of membrane function $\tilde{\eta}$ and a radial component of the magnetic flux density $B_r = \mu_0\left(\mu\frac{\partial \tilde{\eta}_n}{\partial r} + \mu_a\frac{1}{r}\frac{\partial \tilde{\eta}_n}{\partial \theta}\right)$. Here $\mu$ and $\mu_a$ are diagonal component and off-diagonal components of the permeability tensor [18]. With such EM boundary conditions, it becomes evident that the membrane function $\tilde{\eta}$ must not only be continuous and differentiable with respect to a normal to the lateral surface of a disk, but (because of the presence of a gyrotropy term $\mu_a$) be also differentiable with respect to a tangent to this surface. There is evidence of the presence of an azimuthal magnetic field on the border circle with clockwise and counterclockwise rotation asymmetry. In this case, the membrane functions $\tilde{\eta}$ cannot be considered as single-valued functions, and the question arises of the validity of the energy orthogonality relation for MS-wave modes.

To restore the ND boundary conditions and thus the completeness of eigenfunctions $\tilde{\eta}$, we need introducing a certain surface magnetic current $\left(j_s^{(m)}\right)_{top}$ circulating on a lateral surface of the disk. This is a topological current, which compensates the term $\left(i\mu_a\frac{1}{r}\frac{\partial \tilde{\eta}}{\partial \theta}\right)_{r=\mathcal{R}^-}$ in the EM boundary conditions [55, 56, 60]. Evidently, for a given direction of a bias magnetic field (that is, for a given sign of $\mu_a$), there can be two, clockwise and counterclockwise, quantities of the circulating magnetic current. The topological current $\left(j_s^{(m)}\right)_{top}$ is defined by the velocity of an irrotational border flow. This flow is observable via the circulation integral of the gradient $\vec{\nabla}_\theta \delta = \frac{1}{\mathcal{R}}\left(\frac{\partial \delta_\pm}{\partial \theta}\right)_{r=\mathcal{R}} \vec{e}_\theta$, where $\delta_\pm$ is a double-valued edge wave function on contour $\mathcal{L} = 2\pi\mathcal{R}$. On a lateral surface of a quasi-2D ferrite disk, one can distinguish two different functions $\delta_{\pm}$, which are the counterclockwise and clockwise rotating-wave edge functions with respect to a membrane function $\tilde{\eta}$. The spin-half wave-function $\delta_{\pm}$ changes its sign when the regular-coordinate angle $\theta$ is rotated by $2\pi$. As a result, one has the eigenstate spectrum of MDM oscillations with topological phases accumulated by the edge wave function $\delta$. A circulation of gradient $\vec{\nabla}_{\theta}\delta$ along contour $\mathcal{L} = 2\pi\mathcal{R}$ gives a non-zero quantity when an azimuth number is a quantity divisible by $\frac{1}{2}$. A line integral around a singular contour $\mathcal{L}$:

$$\frac{1}{\mathfrak{R}}\oint_{\mathcal{L}}\left(i\frac{\partial \delta_{\pm}}{\partial \theta}\right)(\delta_{\pm})^{*}\,d\mathcal{L} = \int_{0}^{2\pi}\left[\left(i\frac{\partial \delta_{\pm}}{\partial \theta}\right)(\delta_{\pm})^{*}\right]_{r=\mathfrak{R}}d\theta$$

is an observable quantity. Because of the existing the geometrical phase factor on a lateral boundary of a ferrite disk, MDM oscillations are characterized by a pseudo-electric field (the gauge field) $\vec{\epsilon}$. The pseudo-electric field $\vec{\epsilon}$ can be found as $\vec{\epsilon}_{\pm} = -\vec{\nabla} \times (\vec{\Lambda}_{\epsilon}^{(m)})_{\pm}$. The field $\vec{\epsilon}$ is the Berry curvature. The corresponding flux of the gauge field $\vec{\epsilon}$ through a circle of radius $\mathfrak{R}$ is obtained as: $K\int_{S}(\vec{\epsilon})_{\pm}\cdot d\vec{S} = K\oint_{\mathcal{L}}(\vec{\Lambda}_{\epsilon}^{(m)})_{\pm}\cdot d\vec{\mathcal{L}} = K(\Xi^{(e)})_{\pm} = 2\pi q_{\pm}$, where $(\Xi^{(e)})_{\pm}$ are quantized fluxes of pseudo-electric fields, $K$ is the normalization coefficient. Each MDM is quantized to a quantum of an emergent electric flux. There are the positive and negative eigenfluxes. These different-sign fluxes should be nonequivalent to avoid the cancellation. It is evident that while integration of the Berry curvature over the regular-coordinate angle $\theta$ is quantized in units of $2\pi$, integration over the spin-coordinate angle $\theta'$ $\left(\theta' = \frac{1}{2}\theta\right)$ is quantized in units of $\pi$. The physical meaning of coefficient $K$ concerns the property of a flux of a pseudo-electric field. The Berry mechanism provides a microscopic basis for the surface magnetic current at the interface between gyrotropic and nongyrotropic media. Following the spectrum analysis of MDMs in a quasi-2D ferrite disk one obtains pseudo-scalar axion-like fields and edge chiral magnetic currents. The anapole moment for every mode $n$ is calculated as [55, 56, 60].

$$\left(a_{\pm}^{(e)}\right)_{n} \propto \mathcal{R}\int_{0}^{d}\oint_{\mathcal{L}}\left[\left(\vec{j}_{s}^{(m)}(z)\right)_{top}\right]_{n}\cdot d\vec{l}\,dz, \tag{12}$$

where $d$ is the disk thickness. The edge magnetic current $\left(j_{s}^{(m)}\right)_{top}$ is a persistent current appearing due to the mesoscopic effect: the magnitude of such a resonant current becomes appreciable when the parameters of the ferrite disk are reduced to the scale of the dipole-dipole quantum phase coherence length of precessing electrons. At the MDM resonances, one has the "spin-orbit" interaction between precessing magnetic dipoles and a persistent orbital magnetic current. When the frequency of the orbital rotation of MDM resonances in a ferrite disk is close to the ferromagnetic resonance frequency, the precessing magnetic dipoles become strongly correlated and one observers fermionization of the system composed of bosons. There are the macroscopic quantum phenomena related to the collective motion of magnetic dipoles coalescing into the same quantum state, described by a single coherent wavefunction of the condensate. This is a fundamentally distinctive feature from a boson condensate created by small-scale exchange-interaction magnons [61].

It is worth noting that along with the circulation of the surface magnetic current $\left(j_s^{(m)}\right)_{top}$ caused by the edge wave function, there is also the quadratic-form circulation due to this function. For the double-valued edge wave functions $\delta_\pm$, we have the following orthonormality condition on contour $\mathcal{L} = 2\pi \mathcal{R}$ [55]:

$$\mathcal{R} \int_0^{2\pi} \left[ \left( i\mu_a \frac{\partial (\delta_\pm)_n}{\partial \theta} \right) (\delta_\pm^*)_{n'} - (\delta_\pm)_n \left( i\mu_a \frac{\partial (\delta_\pm)_{n'}}{\partial \theta} \right)^* \right]_{r=\mathcal{R}} d\theta$$
$$= \mathcal{R}\mu_a \left[ (q_\pm)_n - (q_\pm)_{n'} \right] \int_0^{2\pi} \left[ (\delta_\pm)_n (\delta_\pm^*)_{n'} \right]_{r=\mathcal{R}} d\theta = 0 \qquad (13)$$

For mode $n$, there are the normalisation relations for the edge functions

$$\int_0^{2\pi} \left[ (\delta_+)_n (\delta_+^*)_n \right]_{r=\mathcal{R}} d\theta = (N_+)_n \qquad (14)$$

and

$$\int_0^{2\pi} \left[ (\delta_-)_n (\delta_-^*)_n \right]_{r=\mathcal{R}} d\theta = (N_-)_n \qquad (15)$$

where $(N_+)_n$ and $(N_-)_n$ are real quantities, which we characterize as surface power flow density of the MDM mode. For a given direction of a bias magnetic field, the wave described by functions $(\delta_+)_n$ propagates only in one direction along the edge. Also, the wave described by function $(\delta_-)_n$ – propagates in one direction (opposite to the former case) along the edge. It is evident that a complex conjugate 'particle' configuration is an 'antiparticle' configuration and vice versa. This fact is related to the existence of only a single edgemode excitation for each wavenumber $q$. In other words, the 'particle' configurations are their own 'antiparticle' configurations. This resembles the well-known properties of the Majorana fermions [62]. So, we have a "chiral Majorana fermion field" [63, 64] on the lateral wall of the MDM ferrite disk.

Due to the presence of surface power flow density, membrane eigenfunction $\tilde{\eta}$ of every MDM rotates around the disk axis. When for every MDM we introduce the notion of an effective mass $\left(m_{lm}^{(eff)}\right)_n$, expressed by Eq. (11), we can assume that for every MDM there exists also an effective moment of inertia $\left(I_z^{(eff)}\right)_n$. With this assumption, an orbital angular momentum a mode is expressed as $(L_z)_n = (I_z^{(eff)})_n \omega$. At the first approximation, let us suppose that the membrane eigenfunction $\tilde{\eta}_n$ is viewed as an infinitely thin homogenous disk of radius $\mathcal{R}$. In other words, we assume that for every MDM, the radial and azimuth variation of the MS-potential function, are averaged. In such a case, we can write

$$\left(I_z^{(eff)}\right)_n = \frac{1}{2}\left(m_{lm}^{(eff)}\right)_n \mathcal{R}^2 d. \tag{16}$$

and

$$\left(L_z^{(eff)}\right)_n = \left(I_z^{(eff)}\right)_n \omega = \frac{\hbar}{4}\beta_n^2 \mathcal{R}^2 d. \tag{17}$$

### C. Spectral problems for MDM magnetostatic oscillations: *L* modes

In the above spectral analysis of the *G* modes, we used the ND boundary conditions. To bridge this spectral problem with the EM boundary conditions, we introduced the contour integrals determining surface magnetic current $\left(j_s^{(m)}\right)_{top}$ and surface power flow density of the MDM modes. However, the MDM spectral can be solved directly based on the EM boundary conditions. This approach is called the *L*-mode spectral analysis. The solution for MS-potential wave function of a *L*-mode is written as

$$\psi_{p,v,q} = C_{p,v,q}\xi_{p,v,q}(z)\tilde{\varphi}_{v,q}(r,\theta), \tag{18}$$

where $C_{p,v,q}$ is a dimensional amplitude coefficient and $\tilde{\varphi}$ is a membrane function. For solutions in a cylindrical coordinate system, one uses the following the boundary condition on a lateral surface of a ferrite disk [52, 53, 55, 56]:

$$\mu\left(\frac{\partial\tilde{\varphi}}{\partial r}\right)_{r=\mathcal{R}^-} - \left(\frac{\partial\tilde{\varphi}}{\partial r}\right)_{r=\mathcal{R}^+} + i\frac{\mu_a}{\mathcal{R}}\left(\frac{\partial\tilde{\varphi}}{\partial \theta}\right)_{r=\mathcal{R}^-} = 0, \tag{19}$$

which are different from the boundary conditions (8). A function $\tilde{\varphi}$ is not a single-valued function. It changes a sign when angle $\theta$ is turned on $2\pi$. For any mode *n*, the function $\tilde{\varphi}_n$ is a two-component sprinor pictorially denoted by two arrows:

$$\tilde{\varphi}_n^{\uparrow\downarrow}(\vec{r},\theta) = \begin{bmatrix}\tilde{\varphi}_n^\uparrow \\ \tilde{\varphi}_n^\downarrow\end{bmatrix} = \tilde{\eta}_n(\vec{r},\theta)\begin{bmatrix}e^{-\frac{1}{2}i\theta} \\ e^{+\frac{1}{2}i\theta}\end{bmatrix} \tag{20}$$

For MS waves in a ferrite medium, described by a *L*-mode scalar wave function $\psi(\vec{r},t)$, we define a magnetic flux density: $\vec{B} = \vec{\vec{\mu}} \cdot \vec{H} = -\vec{\vec{\mu}} \cdot \vec{\nabla}\psi$. In this case, the power flow density can be viewed as a current density, is expressed as [55]:

$$\vec{\mathcal{J}} = \frac{i\omega}{4}\left(\psi \vec{B}^* - \psi^* \vec{B}\right). \tag{21}$$

Such a power flow can appear because of dipole-dipole interaction of magnetic dipoles. With use of separation of variables and taking into account a form of tensor $\vec{\vec{\mu}}$ [18], we decompose a magnetic flux density by two components:

$$\vec{B} = \vec{B}_\perp + \vec{B}_\parallel. \tag{22}$$

The component $\vec{B}_\perp$ are given as

$$\vec{B}_\perp = -\vec{\vec{\mu}}_\perp \cdot \vec{\nabla}_\perp \psi = -C_n \xi_n(z)\left[\vec{\vec{\mu}}_\perp \cdot \vec{\nabla}_\perp \tilde{\varphi}_n(r,\theta)\right]\vec{e}_\perp, \tag{23}$$

where $\vec{e}_\perp$ is a unit vector laying in the $r, \theta$ plane, and

$$\vec{\vec{\mu}}_\perp = \mu_0 \begin{bmatrix} \mu & i\mu_a \\ -i\mu_a & \mu \end{bmatrix}. \tag{24}$$

For the component $\vec{B}_\parallel$ we have

$$\vec{B}_\parallel = -\mu_0 \cdot \vec{\nabla}_\parallel \psi = -\mu_0 C_n \frac{\partial \xi_n(z)}{\partial z} \tilde{\varphi}_n(r,\theta)\vec{e}_z, \tag{25}$$

where $\vec{e}_z$ is a unit vector directed along the *z* axis.

The above representations allow considering two components of the power flow density (current density). For mode *n*, we can write Eq. (21) as

$$\vec{\mathcal{J}}_n = \left(\vec{\mathcal{J}}_\perp\right)_n + \left(\vec{\mathcal{J}}_\parallel\right)_n, \tag{26}$$

where $\left(\vec{\mathcal{J}}_\perp\right)_n = \frac{i\omega}{4}\left[\psi_n\left(\vec{B}_\perp^*\right)_n - \psi_n^*\left(\vec{B}_\perp\right)_n\right]$ and $\left(\vec{\mathcal{J}}_\parallel\right)_n = \frac{i\omega}{4}\left[\psi_n\left(\vec{B}_\parallel^*\right)_n - \psi_n^*\left(\vec{B}_\parallel\right)_n\right]$. Along every of the coordinates $\vec{r}, \vec{\theta}$, and $\vec{z}$, we have the power flows (currents):

$$\left(\vec{\mathcal{J}}_r\right)_n = -\frac{i\omega}{4}|C_n|^2|\xi_n|^2 \mu_0 \left\{\tilde{\varphi}_n\left(\mu\frac{\partial \tilde{\varphi}_n}{\partial r} + i\mu_a \frac{1}{r}\frac{\partial \tilde{\varphi}_n}{\partial \theta}\right)^* - \tilde{\varphi}_n^*\left(\mu\frac{\partial \tilde{\varphi}_n}{\partial r} + i\mu_a \frac{1}{r}\frac{\partial \tilde{\varphi}_n}{\partial \theta}\right)\right\}\vec{e}_r, \tag{27}$$

$$\left(\vec{\mathcal{J}}_\theta\right)_n = -\frac{i\omega}{4}|C_n|^2|\xi_n|^2\mu_0\left\{\tilde{\varphi}_n\left(-i\mu_a\frac{\partial\tilde{\varphi}_n}{\partial r}+\mu\frac{1}{r}\frac{\partial\tilde{\varphi}_n}{\partial\theta}\right)^* - \tilde{\varphi}_n^*\left(-i\mu_a\frac{\partial\tilde{\varphi}_n}{\partial r}+\mu\frac{1}{r}\frac{\partial\tilde{\varphi}_n}{\partial\theta}\right)\right\}\vec{e}_\theta, \tag{28}$$

$$\left(\vec{\mathcal{J}}_z\right)_n = -\frac{i\omega}{4}\mu_0|C_n|^2|\tilde{\varphi}|^2\left[\xi_n\left(\frac{\partial\xi_n}{\partial z}\right)^* - \xi_n^*\frac{\partial\xi_n}{\partial z}\right]\vec{e}_z, \tag{29}$$

where $\vec{e}_r, \vec{e}_\theta$, and $\vec{e}_z$ are the unit vectors. We can see that for membrane function $\tilde{\varphi}$, defined by Eq. (20), there is a non-zero real azimuth component of the power-flow density. So, there is a non-zero quantity of the power flow circulation (clockwise or counterclockwise) around a circle $L = 2\pi r$, where $0 < r \leq \mathcal{R}$. At the same time, homogeneous EM boundary conditions imposed on a ferrite disk on the $r$ and $z$ axes give standing waves without real power flows.

### 3. Near fields of MDM oscillations – the ME near fields

The *L*-mode wave function $\psi(\vec{r},t)$ can define a magnetic flux density in a ferrite disk, as shown above. This scalar wave function is considered as a generating function for other types of the fields both inside and outside a ferrite disk. It allows analyzing complex topological properties of vectorial fields, associated with orbital angular momentum properties of MDM resonances.

When the spectral problem for the MS-potential scalar wave function $\psi(\vec{r},t)$, expressed by Eq. (18), is solved, distribution of magnetization in a ferrite disk is found as $\vec{m} = -\vec{\chi} \cdot \vec{\nabla}\psi$, where $\vec{\chi}$ is the susceptibility tensor of a ferrite [18]. Based on the known magnetization $\vec{m}$, one can find the magnetic field distribution at any point outside a ferrite disk [1, 65]:

$$\vec{H}(\vec{r}) = \frac{1}{4\pi}\left(\int_V \frac{(\vec{\nabla}'\cdot\vec{m}(\vec{r}'))(\vec{r}-\vec{r}')}{|\vec{r}-\vec{r}'|^3}dV' - \int_S \frac{(\vec{n}'\cdot\vec{m}(\vec{r}'))(\vec{r}-\vec{r}')}{|\vec{r}-\vec{r}'|^3}dS'\right). \tag{30}$$

Also, the electric field in any point outside a ferrite disk is defined as [56, 65]

$$\vec{E}(\vec{r}) = -\frac{1}{4\pi}\int_V \frac{\vec{j}^{(m)}(\vec{r}')\times(\vec{r}-\vec{r}')}{|\vec{r}-\vec{r}'|^3}dV', \tag{31}$$

where $\vec{j}^{(m)} = i\omega\mu_0\vec{m}$ is the density of a bulk magnetic current and frequency $\omega$ is the MDM resonance frequency. In Eqs. (30) and (31), $V$ and $S$ are a volume and a surface of a ferrite sample, respectively. Vector $\vec{n}'$ is the outwardly directed normal to surface $S$.

Depending on a direction of a bias magnetic field, we can distinguish the clockwise and counterclockwise topological-phase rotation of the fields. At the MDM resonances, for the magnetic and electric fields defined by Eqs. (30) and (31) one can compose a vector

$$\left\langle \vec{S}_{MDM} \right\rangle^{\uparrow\downarrow} \equiv \frac{1}{2}\text{Re}\left(\vec{E}\times\vec{H}^*\right). \quad (32)$$

The vector $\left\langle \vec{S}_{MDM} \right\rangle$ can be considered as a power flow density vector. Really, based on the vector relation $\vec{\nabla}\cdot\left(\vec{E}\times\vec{H}^*\right) = \vec{H}^*\cdot\vec{\nabla}\times\vec{E}^* - \vec{E}\cdot\vec{\nabla}\times\vec{H}^*$ with taking into account equations $\vec{\nabla}\times\vec{E} = -i\omega\vec{B}$, $\vec{H} = -\vec{\nabla}\psi$ and $\nabla\cdot\vec{B} = 0$, one has as a result $\vec{\nabla}\cdot\left(\vec{E}\times\vec{H}^*\right) = i\omega\vec{\nabla}\cdot\left(\psi^*\vec{B}\right)$. The right-hand side of this equation is a divergence of the power flow density of monochromatic MS waves [55]. So, vector $\left\langle \vec{S}_{MDM} \right\rangle$ can be interpreted as the power flow density as well. Nevertheless, this is not the "EM Poynting vector". Compare to the case of EM wave propagation (with both curl electric and curl magnetic fields), we have here the modes with *curl* electric and *potential* magnetic fields. As we noted above, that there is no EM laws describing transformation of the curl electric field to the potential magnetic field.

In the MDM resonance, the orbital angular-momentum of the power flow density is expressed as

$$\vec{\mathcal{L}}_z = \frac{1}{2}\text{Re}\left[\vec{r}\times\left(\vec{E}\times\vec{H}^*\right)\right]. \quad (33)$$

Depending on a direction of a bias magnetic field, we can distinguish the clockwise and counterclockwise topological-phase rotation of the fields outside the ferrite disk. The direction of an orbital angular-momentum $\vec{\mathcal{L}}_z$ is correlated with the direction of a bias magnetic field $\vec{H}_0$ (along $+z$ axis or $-z$ axis). The active power flow of the field both inside and outside a subwavelength ferrite disk has the vortex topology. In Refs. [65 – 67] it was shown that for every MDM mode, the power flow circulation calculated by Eq. (28) have the same distributions on the $r, \theta$ plane, as the circulation of the power flow vector $\left\langle \vec{S}_{MDM} \right\rangle$. Such the analytically derived distributions coincide with the numerical patterns of the power flows. The analysis was made for a YIG ferrite disk of a 3 mm diameter and thickness of 0.050 mm at the frequency region 8 – 9 GHz. Fig. 1 gives an example of the analytically derived power-flow-density distribution. Fig. 2 shows some numerical patterns of the power flows. One can see a strong confinement of the fields arising from the vortices of the MDM resonances. In Fig. 3, we give a schematic representation of the circulation of the power flow, depicted on the surface of the vacuum sphere and on the surface of the solid angle. Direction of an orbital angular-momentum of a ferrite disk is correlated with the direction of a bias magnetic field.

A persistent edge magnetic current circulating along the contour $\mathcal{L} = 2\pi\mathcal{R}$ on a lateral surface of ferrite disk determines an angular momentum – the anapole moment. Another type of an angular momentum is associated with the power-flow circulation. In a lossless ferrite disk, circulation of the power flow density can be considered as a persistent current as well. The divergentless power-flow-density persistent current, circulating on the $r, \theta$ plane, is an intrinsic property of the fields at the MDM resonances unrelated to the rigid-body rotation of a ferrite-disk. In the Introduction, we asked a question about the possibility of observing a dot product $\vec{E}\cdot\vec{H}$ together with a cross product $\vec{E}\times\vec{H}$ in the near-field region of a subwavelength sample. This question concerned the samples with ME properties. When, for MDM oscillations in a subwavelength ferrite disk, we observe the cross-product of the fields, can we classify this field structure as the ME fields, which are also characterized by the properties of $\mathcal{PT}$-symmetry and dot-product $\vec{E}\cdot\vec{H}$ of the fields? In Ref. [56] it was shown that in the near-field region adjacent to the MDM ferrite disk, there exists also another quadratic parameter determined by the scalar product between the electric and magnetic field components:

$$F = \frac{\varepsilon_0}{4}\operatorname{Im}\left[\vec{E}\cdot\left(\nabla\times\vec{E}\right)^*\right] = \frac{\omega\varepsilon_0\mu_0}{4}\operatorname{Re}\left(\vec{E}\cdot\vec{H}^*\right) \neq 0. \tag{34}$$

This effect is due to the presence of both the curl and potential electric fields in the subwavelength region of the MDM ferrite disk. At the same time, the magnetic near field is pure potential. Parameter $F$ is the ME-field helicity density. It appears only at the MDM resonances. A sign of the helicity parameter depends on a direction of a bias magnetic field. Because of time-reversal symmetry breaking, all the regions with positive helicity become the regions with negative helicity (and vice versa), when one changes a direction of a bias magnetic field:

$$F^{\vec{H}_0\uparrow} = -F^{\vec{H}_0\downarrow}. \tag{35}$$

An integral of the ME-field helicity over an entire near-field vacuum region should be equal to zero [68, 69]. This "helicity neutrality" can be considered as a specific conservation law of helicity. The helicity parameter $F$ is a pseudoscalar: to come back to the initial stage, one has to combine a reflection in a ferrite-disk plane and an opposite (time-reversal) rotation about an axis perpendicular to that plane. The helicity-density distribution is related to the angle between the spinning electric and magnetic fields. Figs. 4 and 5 show the magnetic and electric field distributions on the upper plane of a ferrite disk for the first MDM resonance at different time phases. For such a field structure one can observe both the cross $\vec{E}\times\vec{H}$ and dot $\vec{E}\cdot\vec{H}$ products in the near-field region. The dot-product distributions (the helicity density distributions) are showed in Fig. 6. When one moves from the ferrite surfaces, above or below a ferrite disk, one observes reduction of the field amplitudes and also variation of the angle between spinning electric and magnetic fields. This angle varies from 0° or 180° (near the disk surfaces) to 90° (sufficiently far from a ferrite disk). The "source" of the helicity factor is the pseudoscalar quantity of the magnetization distribution in a ferrite disk at the MDM resonances [68]

$$\operatorname{Im}\int_{V^{(\pm)}}\left[\vec{m}\cdot\left(\vec{\nabla}\times\vec{m}\right)^*\right]dV \neq 0, \tag{36}$$

where $V^{(\pm)}$ are volumes of the upper and lower halves of the ferrite disk. These magnetization parameters are distributed asymmetrically with respect to the z-axis (see Figure 7). Thus, the distribution of the helicity factor is also asymmetric. The regions with nonzero helicity factors we can characterize as the regions with nonzero ME energies. The area with positive helicity factor $F^{(+)}$ is the area with positive ME energy, $W_{ME}^{(+)}$. The area with negative helicity factor $F^{(-)}$ is the area with negative ME energy, $W_{ME}^{(-)}$. The total "ME potential energy" is related to the "ME kinetic energy" of the power-flow rotation. It a symmetrical structure, we have "magnetoelectrically neutral" condensate.

At the MDM resonances, both the power-flow vortices and the helicity states of ME fields are topologically protected quantumlike states. In Ref. [69], it was shown that the power-flow density and the helicity are the complex quantities. In the absence of losses and sources, there exist also the vector $\text{Im}\,\vec{E} \times \vec{H}^*$. This vector can be classified as the reactive power flow density. Fig. 8 illustrates the active and reactive power flows distributions at the MDM resonance above and below a ferrite disk. We can see that while the active power flow is characterized by the vortex topology, the reactive power flow has a source which is originated from a ferrite disk. The regions of localization of the active and reactive power flows are different. While the active power flow is localized at the disk periphery, the reactive power flow is localized at a central part of the disk. It was shown [69] that above and below a quasi-2D ferrite disk, the real part of the helicity density (defined by Eq. (34)) is related to an imaginary part of the complex power-flow density:

$$\frac{1}{2}\text{Re}\left|\vec{E}\cdot\vec{H}^*\right| = \frac{1}{2}\text{Im}\left|\left[\vec{E}\times\vec{H}^*\right]_z\right| \tag{37}$$

The numerical results in Ref. [69] clearly show that in a vacuum region where the helicity-density factor exist, the reactive power flow is observed as well. So, in a region near a ferrite disk, the reactive power flow is accompanied by the helicity factor or, in other words, by the ME-energy density.

The pseudoscalar parameter (36) and the helicity factor $F$, arise due to spin-orbit interaction. Such *PT*-symmetric parameters, mixing electric and magnetic fields, are associated with the axion-electrodynamic term, leading to modification of inhomogeneous Maxwell equations [57, 70, 71]. It means that the ME fields appear as the fields of axion electrodynamics. With such a unique topological structure of ME near fields, two types of polaritons should be observed: right-handed ME polaritons and left-handed ME polaritons. When an external microwave structure is geometrically symmetrical, the two types of ME polaritons are indistinguishable. Otherwise, different microwave responses could be observed depending on the direction of the bias magnetic field. When an external microwave structure contains any elements with geometrical chirality, the right-hand and left-hand ME polaritons becomes nondegenerate, and microwave responses depend on the direction of the bias magnetic field. This fact was confirmed both numerically and experimentally [56].

## 4. MDM particles inside waveguides and cavities.

In microwaves, we are witnesses that long-standing research in coupling between electrodynamics and magnetization dynamics noticeably reappear in recent studies of strong magnon-photon interaction [72 – 76]. In a small ferromagnetic particle, the exchange interaction can lead to the fact that a very large number of spins to lock together into one macrospin with a corresponding increase in oscillator strength. This results in strong enhancement of spin-photon coupling. In a structure of a microwave cavity with a yttrium iron garnet (YIG) sphere inside, the avoided crossing in the microwave reflection spectra verifies the strong coupling between the microwave photon and the macrospin magnon. In these studies, the Zeeman energy is defined by a coherent state of the macrospin-photon system when a magnetic dipole is in its antiparallel orientation to the cavity magnetic field. Together with an analysis of the strong coupling of the electromagnetic modes of a cavity with the fundamental Kittel modes, coupling with non-uniform modes – the Walker modes – in a YIG sphere was considered. In the microwave experiments, identification of the Walker modes in the sphere was made based an effect of overlapping between the cavity and spin waves due to relative symmetries of the fields [77, 78]. Nevertheless, the experimentally observed effects of strong magnon-photon interaction, cannot be described properly in terms of a single magnon-photon coupling process. In a view of these aspects, the theory based on solving coupled Maxwell and Landau-Lifshitz-Gilbert equations without making the conventional magnetostatic approximation have been suggested [79, 80]. Currently, the studies of strong magnon-photon interaction are integrated in a new field of research called cavity spintronics (or spin cavitronics) [81].

The coupling strength in the magnon-photon system is proportional to the probability of conversion of a photon to a magnon and vice versa. An effective way for strong coupling is to confine both magnons and photons to a small (subwavelength) region. Long-range spin transport in magnetic insulators demonstrates that the dipolar interactions alone generate coherent spin waves on the scales that are much larger than the exchange-interaction scales and, at the same time, much smaller than the electromagnetic-wave scales. Because of symmetry breakings, the MDM ferrite disk, being a very small particle compared to the free-space electromagnetic wavelength, is a singular point for electromagnetic fields in a waveguide or cavity. When we consider a ferrite disk in vacuum environment, the unidirectional power-flow circulation might seem to violate the law of conservation of an angular momentum in a mechanically stationary system. In a microwave structure with an embedded ferrite disk, an orbital angular momentum, related to the power-flow circulation, must be conserved in the process. It can be conserved if topological properties of electromagnetic fields in the entire microwave structure are taken into account. Thus, if power-flow circulation is pushed in one direction in a ferrite disk, then the power-flow circulation on metal walls of the waveguide or cavity to be pushed in the opposite direction at the same time. It means that, in a general consideration, the model of MDM-vortex polaritons appears as an integrodifferential problem. Fig. 9 presents a schematic picture of an interaction of a MDM ferrite disk with an external microwave structure. In Ref. [68] it was shown that due to the topological action of the azimuthally unidirectional transport of energy in a MDM-resonance ferrite sample there exists the opposite topological reaction (opposite azimuthally unidirectional transport of energy) on a metal screen placed near this sample. It is obvious that the question of the interaction of a MDM ferrite disk with an external microwave structure is far from trivial. To illustrate this nontriviality in more details, we adduce here some topological problems related to our studies.

## A. On Rayleigh scattering by a thin ferrite rod

In the above studies of the MDM oscillation spectra in an open quasi-2D ferrite disk, the separation of variables in a cylindrical coordinate system was used. Analytically, we cannot apply a 2D model to consider scattering of EM waves by a subwavelength ferrite disk. Nevertheless, based on a simple qualitative analysis of a 2D structure, we can illustrate the role of topology in the EM-wave scattering by a ferrite sample. For this purpose, we will view some properties of the EM-wave scattering by a thin endless ferrite rod in comparison with the EM-wave scattering by a thin endless metal rod.

In Fig. 10, we give a schematic illustration of charges and currents on the cross-section of the rods at the dipole-like scattering. Let us consider, initially, Rayleigh scattering by a thin endless cylindrical rod made from a perfect electric conductor (PEC). A rod oriented along the $z$ axis is acted upon by an external alternating electric field in the plane $r, \theta$ of a plane electromagnetic wave. Assuming that the rod diameter is much less than the EM wavelength, the analysis can be viewed as a quasi-electrostatic problem. The electric field of the EM wave induces positive and negative electric charges on diametrically opposite points of the $r, \theta$ plane, which cause two, clockwise (CW) and counter clockwise (CCW), azimuthal electric currents on the rod surface. Creating an azimuthally symmetric structure, each of these surface currents passes over a regular-coordinate angle $\pi$. In such a structure we have both the azimuthal and time symmetries. In Fig. 10 (*a*), the $+,-$ surface electric charges correspond to the maximum, minimum of the charge distributions in the azimuth coordinates. One can adduce other examples of the azimuthal and time symmetries (the *PT* symmetry) at the dipole-like scattering from subwavelength structures. This includes also the electric-dipole eigenmodes of the surface plasmon resonances [82 – 84].

We consider now a thin endless cylindrical rod made from a magnetic insulator, YIG. A plane EM wave propagates along the rod axis. The rod diameter is much less than the EM wavelength and the analysis is considered as a quasi-magnetostatic problem. The rod is axially magnetized up to saturation by a bias magnetic field directed along the *z* axis. Due to the anisotropy (gyrotropy) induced by bias magnetic field, the RF magnetic field of the EM wave, which lies in the $r,\theta$ plane, causes the precessional motion of the alternating magnetization vector about the *z* axis. In this structure, magnetic charges at diametrically opposite points of a ferrite rod can appear due to the divergence of magnetization. It is known that at the ferromagnetic resonance frequency in an infinite medium, no divergence of the magnetization exists. Also, it is known that a divergence of the DC magnetization exists in a ferrite ellipsoid (in an endless cylindrical ferrite rod, in particular) with the homogeneous-precession mode (Kittel's mode) [18]. The divergence of both the DC and RF magnetizations may occur in a ferrite sample in a case of nonhomogeneous-precession modes. These magnetic-dipole modes – Walker's modes – in a thin endless cylindrical ferrite rod were studied in Ref. [85]. For these eigenmodes, the RF magnetic field of the incident EM wave induces a magnetic dipoles, which lies in the $r,\theta$ plane of a ferrite rod. In Fig. 10 (*b*), the $+^{(m)},-^{(m)}$ surface magnetic charges correspond to the maximum, minimum of the charge distributions in the azimuth coordinates. Due to the time-reversal symmetry breaking, these surface magnetic charges cannot cause two, clockwise (CW) *and* counter clockwise (CCW), azimuthal magnetic currents. For the given direction of bias magnetic field, we may have only CW *or* CCW induced magnetic current, which passes over a regular-coordinate angle $\pi$ at the time phase of $\pi$. In Fig. 10 (*b*), this is shown as the CW blue-arrow current. We can suppose, however, that there exists a polaritonic structure with an additional (non-electromagnetic) phase shift, when the gradient of twisting angle plays the role of the phase gradient. A global phase texture with coflowing an EM-wave induced magnetic current and topological magnetic current will provide us with the possibility to have rotational symmetry by a turn over a regular-coordinate angle $2\pi$ at the time phase of $\pi$. This situation is shown in Fig. 10 (*b*), where the CW magnetic current of a polaritonic structure is conventionally represented as a circle composed by the blue and red arrows. It is worth noting, however, that one can view such a phenomenon not in a ferrite rod, but in the $r,\theta$ plane of a ferrite disk with MDM oscillations, where the non-electromagnetic torque is caused by the topological-phase effects. In the MDM ferrite-disk resonator, the non-zero circulation of such a magnetic current, observed at the time phase shift of $\pi$, results in appearance of a constant angular momentum directed along the *z* axis. This is possible due to an additional phase shift of the magnetic current along the *z* axis. The magnetic currents have a helical structure. When such helical currents (and so helical waves) cannot be observed in a smooth ferrite rod, they can be seen in a MDM ferrite disk [86].

**B. Testing the topological properties of the ME field with small metal rods and rings in a microwave waveguide**

In a structure of a MDM particle embedded in a microwave waveguide, photons interact strongly and coherently with magnetic excitations. The creation of certain non-classical states in such a macroscopic system can be observed with help of small metallic elements placed inside a microwave waveguide near the ferrite disk. Here we show some of the topological properties of the ME field using small metallic rods and rings.

A structure of microwave waveguide with a ferrite disk and small metallic rod, shown in Fig. 11 (*a*), was studied experimentally in Ref. [87] and numerically in Ref. [88]. The rod is oriented along an electric field of a rectangular waveguide. Its diameter is a very small compared to the disk diameter and to the free-space electromagnetic wavelength. On the basis of a comparative analysis of experimental oscillation spectra, it was argued in Ref. [87] that the fact that an additional small capacitive coupling (due to a piece of a nonmagnetic wire) strongly affects magnetic oscillation proves the presence of the electric-dipole moments (anapole moments) of the MDMs in a quasi-2D ferrite disk. In numerical studies [88], a metal rod is made of a PEC. At frequencies far from the MDM resonances, the field structure of an entire waveguide is not noticeably disturbed. The electric field on the rod demonstrates a trivial picture of the field induced on a small electric dipole inside a waveguide [Fig. 11 (*b*)]. At the same time, in the case of the MDM resonance, there is a strong reflection of electromagnetic waves in a waveguide. The PEC rod behaves as a small line defect on which rotational symmetry is violated. The observed evolution of the radial part of the electric polarization, giving, as a result, a circulating electric current, indicates the presence of a geometrical phase in the vacuum-region field of the MDM-vortex polariton [Fig. 11 (*c*)].

Let us bend the metallic rod into a ring and rigidly connect the ends. At the MDM resonance, rotating electric charges and circulating electric currents arise on a ring placed above the ferrite disk. [89]. Fig.12 shows circulation of a surface electric current along a PEC ring. The ferrite disk has a diameter of 3 mm. A metallic ring made from a wire of a diameter of 0,05 mm, has a diameter of 1.5 mm. The ring is located above a ferrite disk at a distance of 0.05 mm. The circulating current will give an angular-momentum flux. The intensity of the flux is proportional to the gradient of twisting angle, which plays the role of the phase gradient. A critical phase gradient is required to enable the process. This occurs only at the MDM resonance. In this case, the persistent charge current in the ring is correlated to the persistent magnetic current in a ferrite disk. The electric current on the surface of the metallic ring has the spin degree of freedom [see Fig. 12 (*c*)]. When using 2D models in our main studies of MDM oscillations, we can conclude now that, generally, an analysis of ME fields should be made based on the 3D model. At the MDM resonance, the current induces on a test metal ring is a topological soliton structure which is quantized simultaneously in poloidal and toroidal directions. This 3D continuous vector-field structure – a hopfion (or Hopf soliton) – cannot be unknotted without cutting [90, 91]. It is also worth noting that the helicity properties of the 3D structure of the ME field in vacuum reflect its own topologically nontrivial structure at *each mode* of the MDM-oscillation spectrum.

It is worth noting that in a remarkable paper [92], the authors had measured the low-temperature magnetization response of an isolated mescoscopic copper ring to a slowly varying magnetic flux. They showed that the total magnetization response oscillates as a function of the enclosed magnetic flux on the scale of *half a flux quantum*. In our study, a numerical analysis made in Ref. [89] shows that the currents induced on a metal ring at the MDM resonances, strongly perturb the electric, but not the magnetic, field in a vacuum region above the ferrite disk. This means that the ring is threaded mainly by an electric flux. Taking into account the above-analyzed properties of the anapole, we have evidence of the presence of the enclosed *electric* flux on the scale of *half a flux quantum*.

## C. MDM cavity electrodynamics

For the case of MDM resonances in a small ferrite disk, characterizing by non-uniform magnetization dynamics, the above-mentioned model of coherent states of the macrospin/photon system in a ferrite sphere [72 – 76, 81], is not applicable. In Refs. [57, 93], it was shown that multiresonance microwave oscillations observed in experiments [49 – 51, 93], are related to the fact that magnetization dynamics of MDM oscillations in a quasi-2D ferrite disk have a strong impact on the phenomena associated with the quantized energy fluctuation of microwave photons in a cavity. Fig. 13 gives a sketch showing the relationship between quantized states of microwave energy in a cavity and magnetic energy in a MDM ferrite disk. The microwave structure is a rectangular waveguide cavity with a normally magnetized ferrite-disk sample. The operating frequency, which is a resonant frequency of the cavity, is constant. The only external parameter, which varies in the experiment, is a bias magnetic field. The observed discrete variation of the cavity impedances is related to discrete states of the cavity fields. Since the effect was obtained at a given resonant frequency, the shown resonances are *not conventional cavity modes* related to the frequency-dependent quantization of the photon wave vector. These resonances are caused by the quantized variation of energy of a ferrite disk, which appear due to variation of energy of an external source – the bias magnetic field.

At the regions of a bias magnetic field, designated in Fig. 13 as *A, a, b, c, d, …*, we do not have MDM resonances. In these regions, a ferrite disk is "seen" by electromagnetic waves, as a very small obstacle which, practically, does not perturb a microwave cavity. In this case, the cavity (with an embedded ferrite disk) has good impedance matching with an external waveguiding structure and a microwave energy accumulated in a cavity is at a certain maximal level. At the MDM resonances, the reflection coefficient sharply increases (the states designated in Fig. 13 by numbers 1, 2, 3, …). The input impedances are real, but the cavity is strongly mismatched with an external waveguiding structure. It means that at the MDM-resonance peaks, the cavity receives less energy from an external microwave source. In these states of a bias magnetic field, the microwave energy stored in a cavity sharply decreases, compared to its maximal level in the *A, a, b, c, d, …* Since the only external parameter, which varies in this experiment, is a bias magnetic field, such a sharp release of the microwave energy accumulated in a cavity to an external waveguiding structure should be related to the emission of discrete portions of energy from a ferrite disk. This means that at the MDM resonances, a strong and sharp decrease in the magnetic energy of the ferrite sample should be observed.

When speaking about the eigenstates of the microwave-cavity spectrum observed at the bias-field variation and constant frequency, we should answer the question about the eigenfunctions of this spectrum. In general, microwave resonators with the time-reversal symmetry breakings give an example of a nonintegrable, i.e., path dependent, system. The time-reversal symmetry breaking effect leads to creation of the Poynting-vector vortices in a vacuum region the microwave resonators with enclosed lossless ferrite samples [94 – 97]. In an analysis of the cavity eigenfunctions, it makes no sense to consider the reflection of electromagnetic waves from magnetized ferrites from the standpoint of energy flow and ray propagation [98]. One cannot use an interpretation which allows viewing the modes as pairs of two bouncing electromagnetic plane waves. This interpretation clearly shows that for a structure with an enclosed magnetized ferrite sample given, for example, in Fig. 14, there can be no identity between the rays $1 \to F \to 1'$ and $1' \to F \to 1$ in the sense that these rays can acquire different phases when are reflected by the ferrite.

At the same time, it is argued [99] that in quantum mechanics the distinction between integrable and nonintegrable systems does not work any longer. The initial conditions are defined only within the limits of the uncertainty relation $\Delta x \Delta p \geq \frac{1}{2} \hbar$. Since the Schrödinger equation is linear, a quantum mechanical wave packet can be constructed from the eigenfunctions by the superposition principle. What do we have in our structure of a MDM ferrite disk in a cavity? We use the Walker equation for a MS scalar wave function. It also allows to construct a wave packet from the eigenfunctions by the superposition principle. We have to use a description of the spectral response functions of the system with respect to two external parameters – a bias magnetic field $H_0$ and a signal frequency $\omega$ – and analyze the correlations between the spectral response functions at different values of these external parameters. It means that, in neglect of losses, there should exist a certain *uncertainty limit* stating that

$$\Delta f \Delta H_0 \geq \text{uncertainty limit}. \tag{38}$$

This uncertainty limit is a constant which depends on the disk size parameters and the ferrite material property (such as saturation magnetization) [57]. Beyond the frames of the uncertainty limit (38) one has continuum of energy. The fact that there are different mechanisms of quantization allows to conclude that for MDM oscillations in a quasi-2D ferrite disk both discrete energy eigenstate and a continuum of energy can exist. In quantum mechanics, the uncertainty principle says that the values of a pair of canonically conjugate observables cannot both be precisely determined in any quantum state. In a formal harmonic analysis in classical physics, the uncertainty principle can be summed up as follows: A nonzero function and its Fourier transform cannot be sharply localized. This principle states also that there exist limitations in performing measurements on a system without disturbing it. Basically, formulation of the main statement of the MDM-oscillation theory is impossible without using a classical microwave structure. If a MDM particle is under interaction with a "classical electrodynamics" object, the states of this classical object change. The character and value of these changes depend on the MDM quantized states and so can serve as its qualitative characteristics. The microwave measurement reflects interaction between a microwave cavity and a MDM particle. It is worth noting that for different types of subwavelength particles (ferrite disks), the uncertainty limits may be different.

The fact of the existence of the uncertainty limit (38) is indirectly confirmed by the experimental results presented in Ref. [100]. In the microwave structure shown in Fig. 15 (*a*), a ferrite disk is placed in a cavity with a very low Q factor. The wide bandwidth is due to losses caused by the test samples embedded in the cavity. Fig. 15 (*b*) shows how a bias magnetic field tunes the shape of the MDM resonance. It can be seen that as one approaches the top of the cavity resonance curve, the effect of Fano resonance collapses, the Fano line shape is completely decays, and a single Lorentz peak is observed. The Lorentzian response is a narrow, highly symmetric peak. The scattering cross section corresponds to a pure dark mode. All this means that, within uncertainty limit (38), it is possible to carry out observations for a very wide linewidth of the cavity mode ($\Delta f$ is very big) and an extremely narrow linewidth of the MDM resonance peak ($\Delta H_0$ is very small).

In the above studies, we considered the *G* modes (with a scalar MS membrane function $\tilde{\eta}$ and the ND BC) and the *L* modes (with vector MS membrane function $\tilde{V} \equiv \begin{pmatrix} \tilde{\vec{B}} \\ \tilde{\varphi} \end{pmatrix}$ and the EM BC). The *G*-mode spectral analysis is more appropriate to use at the regime of a constant frequency and the bias magnetic-field variation, while the *L*-mode analysis – at a constant bias magnetic field and the frequency variation. These two spectral problems are bridged within uncertainty limit (38).

## 5. Transfer of angular momentum to dielectric materials, metals and biological structures from MDM resonators

Due to unique structures of twisted ME near fields, one can observe angular momentums (spin and orbital) transfer to electric polarization in a dielectric sample (Fig. 16). Experiments [101, 102] show explicit shifts of the MDM resonance peaks due to the dielectric loading of the ferrite disk. This effect was explained in Refs. [56, 57, 102]. The mechanical torque exerted on a given electric dipole in a dielectric sample is defined as a cross product of the MDM electric field and the electric moment of the dipole. The torque exerting on the electric polarization in a dielectric sample due to the MDM electric field should be equal to reaction torque exerting on the magnetization in a ferrite disk. Because of this reaction torque, the precessing magnetic moment density of the ferromagnet will be under additional mechanical rotation at a certain frequency $\Omega$. The frequency $\Omega$ is defined based on both, spin and orbital, momentums of the fields of MDM oscillations. It was shown experimentally that the chiral structure of near-field ME provides the potential for microwave chirality discrimination in chemical and biological objects [103].

Because of a chiral topology of near fields originated from MDM oscillations in a ferrite disk, one has helical electric currents induced on a surface of a metal wire electrode placed on a surface of a ferrite disk. On a butt end of a wire probe one can observe twisted near fields (Fig. 17). The handedness of these fields depends on a direction of a bias magnetic field applied to a MDM ferrite resonator [102]. Using helical electric currents induced on a metal wire electrode, one can obtain the angular momentum transfer to localized regions in dielectric samples (Fig. 18).

Due to strong reflection and absorption of electromagnetic waves in conductive layers and biological tissue, standard microwave techniques cannot be used for testing such structures. Twisted microwave near fields with strong energy concentration, originated from MDM ferrite disk with a metal wire electrode, allow probing effectively high absorption conductive layers. This effect can be explained by a simple physical model. When the electromagnetic wave incidents on a conductive material, the induced electric current is almost parallel to the electric field (Ohm's law). Joule losses in conductive materials are defined by a scalar product of an induced electric current and an electric field. When, however, a conductive material is placed in a twisted microwave near field, the RF electric current and an RF electric field become mutually nonparallel. It means that for Joule losses, one has $\vec{J} \cdot \vec{E} = J \cdot E \cos \delta$ with $\cos \delta \ll 1$. Extremely small Joule losses result in strong enlargement of a penetration length – the skin depth – in a sample. Fig. 19 presents numerical results illustrating the effect of penetration of the twisted-field microwave power through a thin metal screen [104].

## 6. Conclusion

ME fields are subwavelength-domain fields with specific properties of violation of spatial and temporal inversion symmetry. When searching for such fields, we consider near fields originated from subwavelength resonators, that are the systems with quantum-confinement effects of dipolar-mode quasistatic oscillations. We show that the near fields of a quasi-2D subwavelength-size ferrite disk with magnetic-dipolar-mode (MDM) oscillations have the properties of ME fields. The ME fields, being originated from magnetization dynamics at MDM resonances, appear as the pseudoscalar axionlike fields. Whenever the pseudoscalar axionlike fields, is introduced in the electromagnetic theory, the dual symmetry is spontaneously and explicitly broken. This results in non-trivial coupling between pseudoscalar quasistatic ME fields and the EM fields in microwave structures with an embedded MDM ferrite disk.

Long range magnetic dipole-dipole correlation can be treated in terms of collective magnetostatic excitations of the system. In small ferromagnetic-resonance ferrite disk, macroscopic quantum coherence can be observed. In a case of a quasi-2D ferrite disk, the quantized forms of these collective matter oscillations – the MDM magnons – were found to be quasiparticles with both wave-like and particle-like behaviors, as expected for quantum excitations. With use of MS-potential scalar wave function $\psi$ we formulate properly the energy eigenstate problem based on the Schrödinger-like equation. We obtain currents (fluxes) for MS modes. We show that in a subwavelength ferrite-disk particle one can observe an angular momentum due to the power-flow circulation of double-valued edge MS-wave functions. For incident electromagnetic wave, this magnon subwavelength particle emerges as a singular point carrying quanta of angular momenta. In a ferrite-disk sample, the magnetization has both the spin and orbital rotations. There is the spin-orbit interaction between these angular momenta. The MDMs are characterized by the pseudoscalar magnetization helicity parameter, which can be considered as a certain source of the helicity properties of ME fields.

Quantized ME fields arising from nonhomogeneous ferromagnetic resonances with spin-orbit effect, suggest a conceptually new microwave functionality for material characterization. Due to unique structures of twisted ME near fields, one can observe angular momentums (spin and orbital) transfer to electric polarization in a dielectric sample. The chiral structure of near-field ME provides the potential for microwave chirality discrimination in chemical and biological objects. Twisted ME fields allow deep penetration of the microwave power into materials with high conductivity.

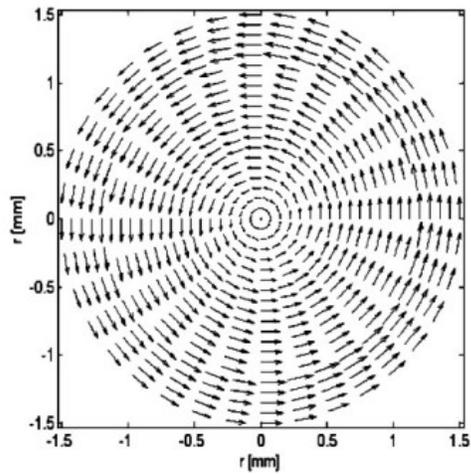

Fig 1. Analytically derived power-flow-density distribution for the main MDM inside a ferrite disk of dimeter 3 mm (arbitrary units).

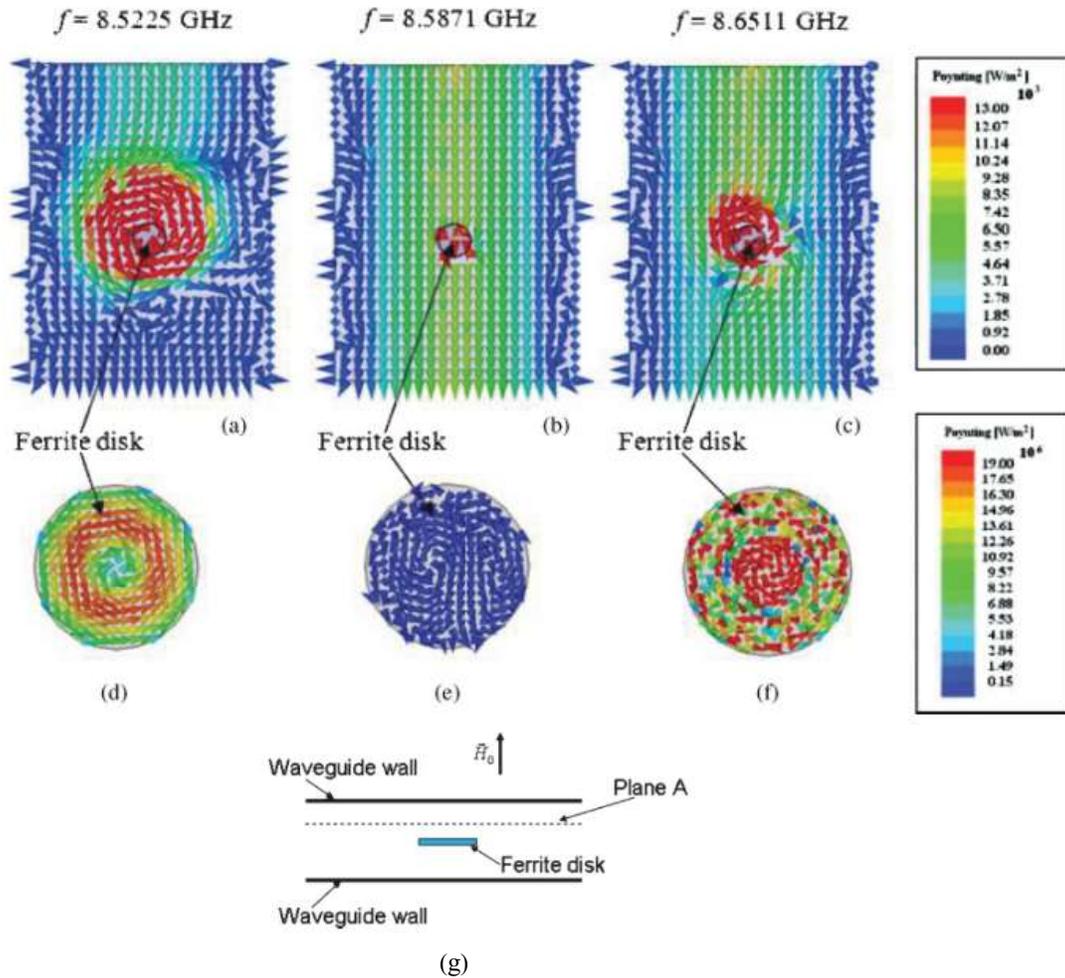

(g)

Fig. 2. Field confinement originating from the MDM vortices in a ferrite disk. (a) The Poynting vector distributions for the field on plane A at the frequency (f = 8.5225 GHz) of the first resonance. (b) The same at the frequency (f = 8.5871 GHz) between the resonances. (c) The same at the frequency (f = 8.6511 GHz) of the second resonance. (d) The Poynting vector distributions inside a ferrite disk at the frequency of the first resonance. (e) The same at the frequency between resonances. (f) The same at the frequency of the second resonance. (g) The plane A is a vacuum plane inside a waveguide situated above an upper plane of a MDM ferrite disk. Ref. [67].

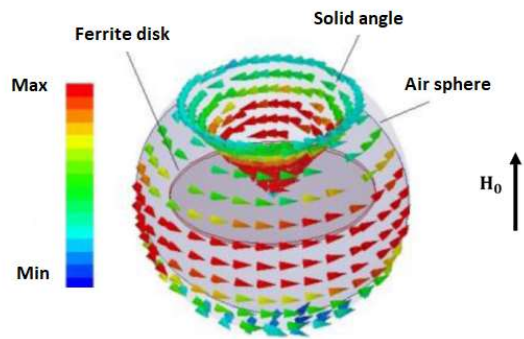

Fig. 3. Schematic representation of the circulation of the power flow, depicted on the surface of the vacuum sphere and on the surface of the solid angle. Direction of an orbital angular-momentum of a ferrite disk is correlated with the direction of a bias magnetic field.

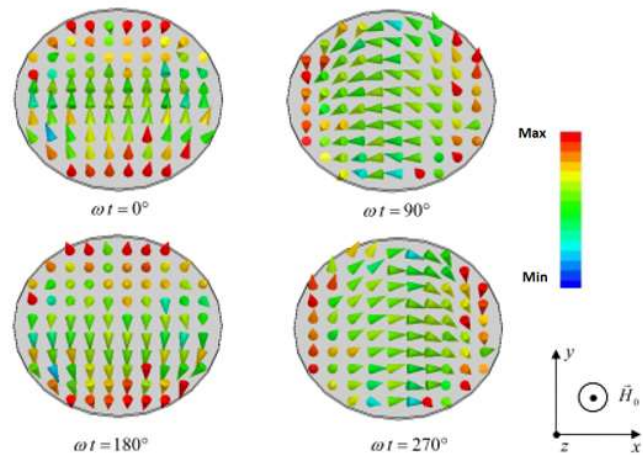

Fig. 4. Magnetic field distributions on the upper plane of a ferrite disk for the first MDM resonance at different time phases.

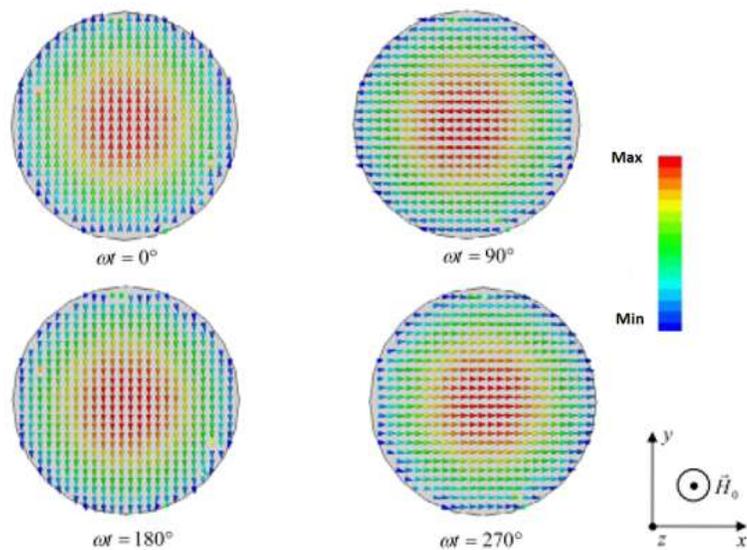

Fig. 5. Electric field distributions on the upper plane of a ferrite disc for the first MDM resonance at different time phases.

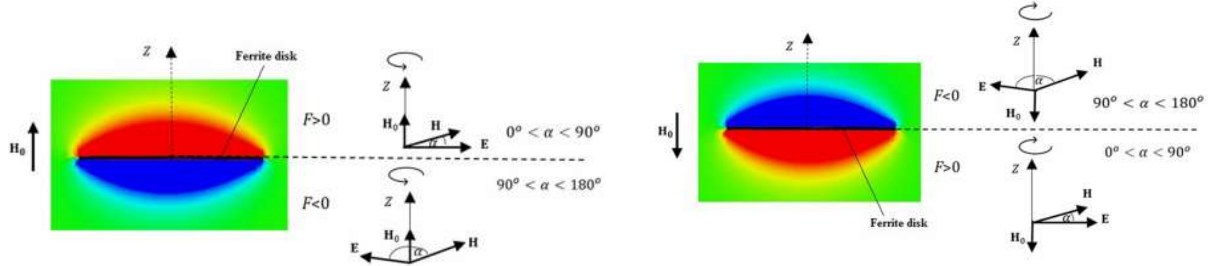

Fig. 6. The helicity density distributions above and below a ferrite disk at the MDM resonance at two opposite directions of a bias magnetic field. The electric and magnetic fields outside a ferrite disk are rotating fields which are not mutually perpendicular. The helicity parameter $F$ is a pseudoscalar: to come back to the initial stage, one has to combine a reflection in a ferrite-disk plane and an opposite (time-reversal) rotation about an axis perpendicular to that plane. In a green region $F = 0$: the angle between the electric and magnetic fields is $90°$.

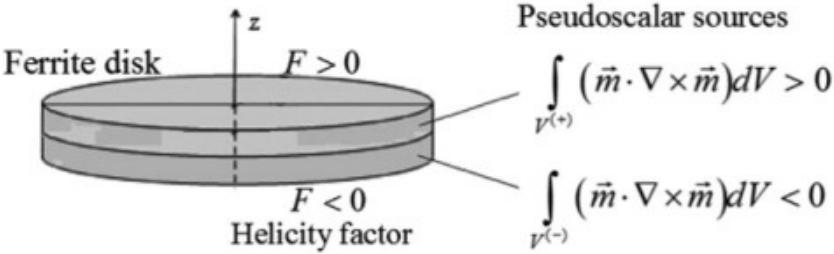

Fig. 7. Pseudoscalar quantity of the magnetization in a ferrite disk as a "source" of the helicity factor at the MDM resonance.

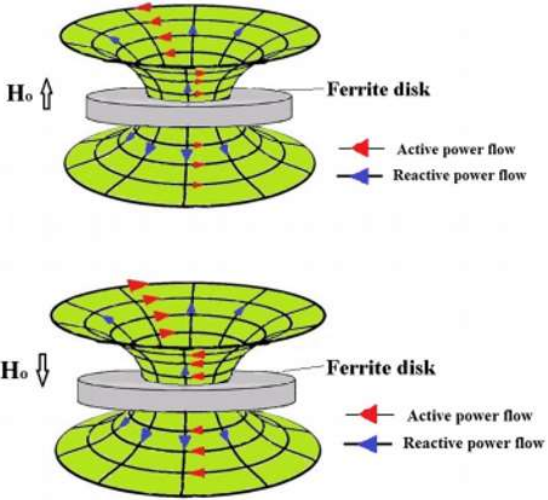

Fig. 8. The active and reactive power flows of the ME field at the MDM resonance. a) An upward directed bias magnetic field; b) a downward directed bias magnetic field. The active and reactive power flows are mutually perpendicular. These flows constitute surfaces, which can be considered as deformed versions of the complex planes, i.e., as Riemann surfaces. When one changes a direction of a bias field, the active power flow changes its direction as well. At the same time, the reactive power flow does not change its direction when the direction of a bias field is changed.

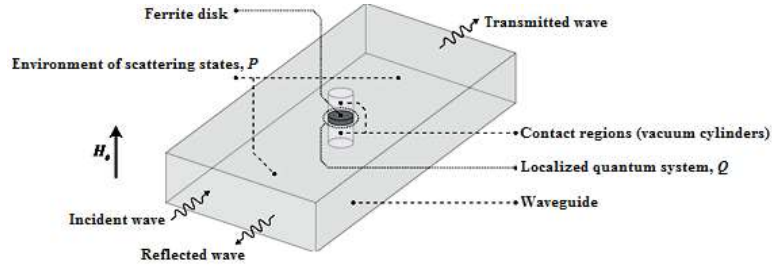

Fig. 9. An interaction of a MDM ferrite disk with a microwave waveguide. The structure is viewed as the $P + Q$ space. It consists of a localized quantum system (the MDM ferrite disk), denoted as the region $Q$, which is embedded within an environment of scattering states (the microwave waveguide), denoted as the regions $P$. The coupling between the regions $Q$ and $P$ is regulated by means of the two "contact regions" in the waveguide space.

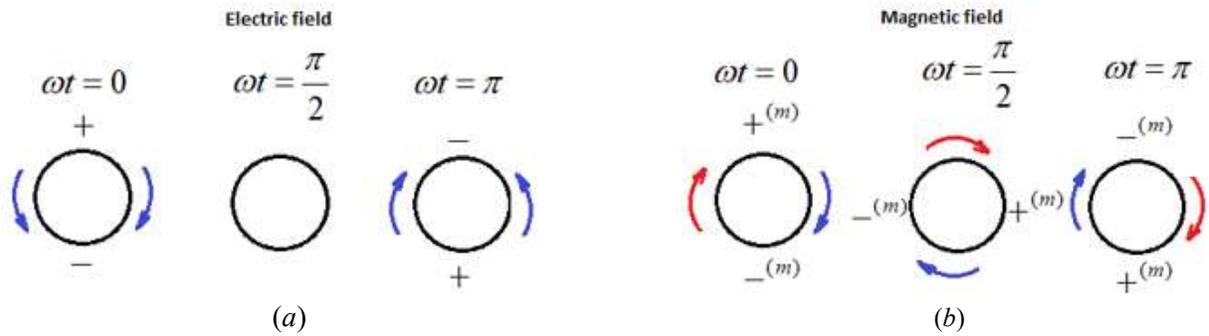

Fig. 10. Schematic illustration of charges and currents on the cross-section of the rods at the dipole-like scattering. (a) Electric charges and currents on a surface of a thin metal rod induced by RF electric field in a $r, \theta$ plane. (b) Magnetic charges and currents on a surface of a thin ferrite rod induced by RF magnetic field in a $r, \theta$ plane. Magnetic current of a polaritonic structure is conventionally represented as a circle composed by the blue and red arrows. In this case, the entire cycle of rotation corresponds to the $\pi$-shift of a dynamic phase.

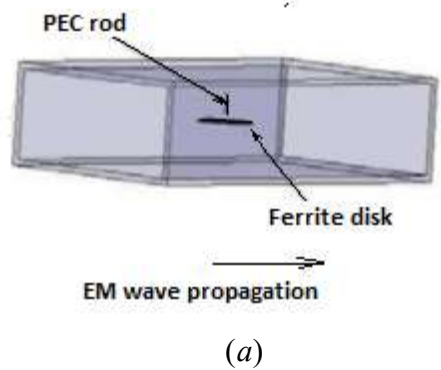

(a)

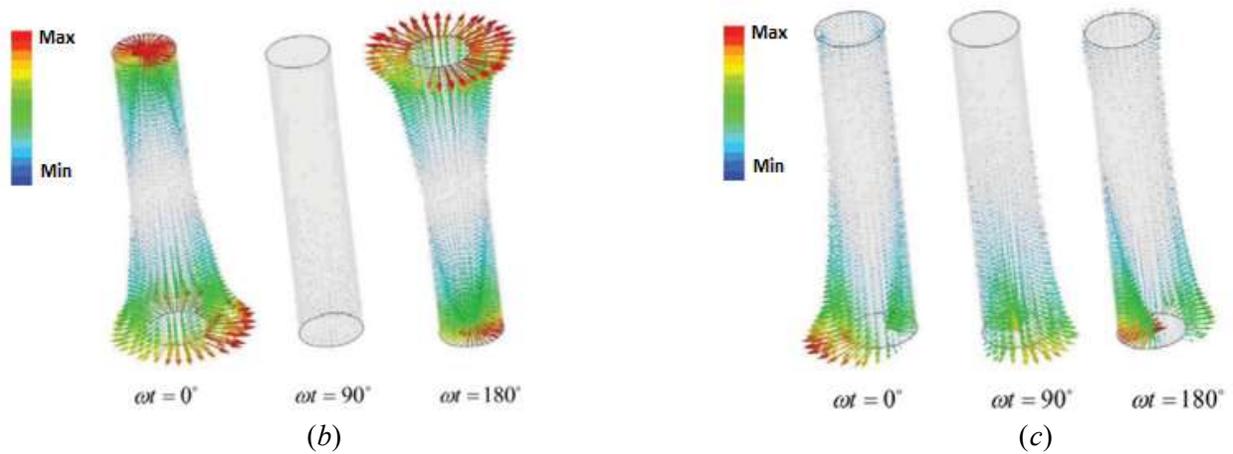

(b)    (c)

Fig. 11. (*a*) A structure of microwave waveguide with a ferrite disk and small metallic rod. (*b*) Electric field on a small PEC rod for the frequency far from the MDM resonance at different time phases. There is a trivial picture of the fields of a small electric dipole inside a waveguide. (*c*) Electric field on a small PEC rod in the MDM resonance at different time phases. A PEC rod behaves as a small line defect on which rotational symmetry is violated. The observed evolution of the radial part of polarization gives evidence for the presence of a geometrical phase in the vacuum-region field of the MDM-vortex polariton.

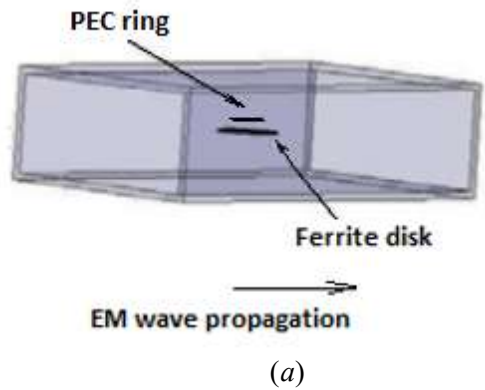

(a)

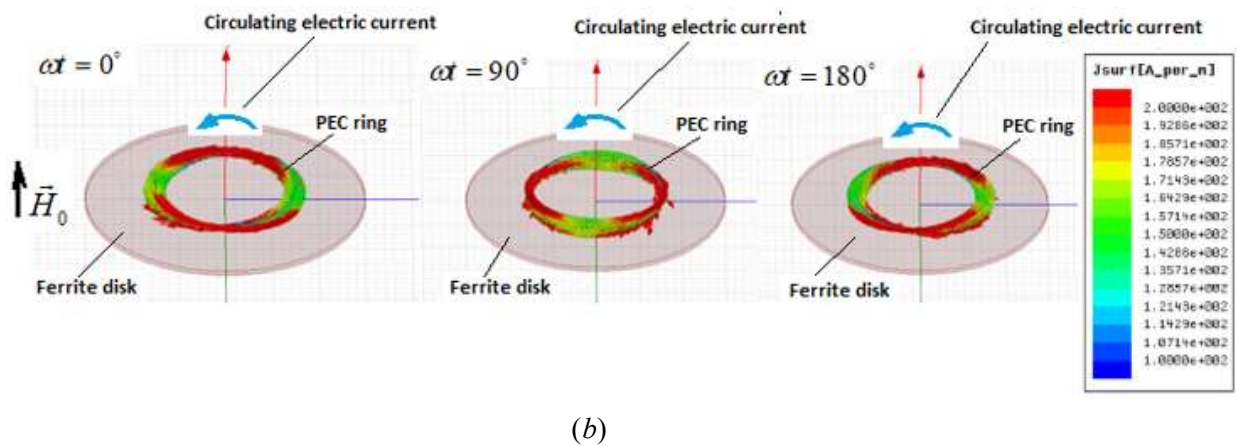

(b)

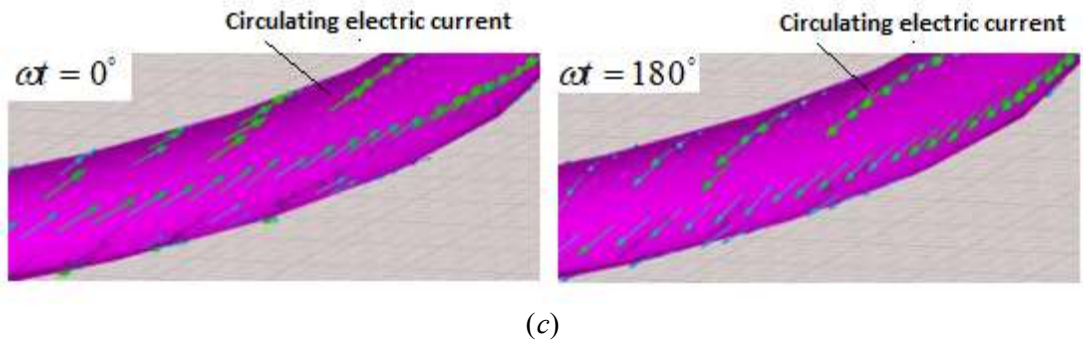

(c)

Fig. 12. (*a*) A structure of microwave waveguide with a ferrite disk and small metallic ring. (*b*) Circulating surface electric current on a metallic ring. (*c*) The electric current on the surface of a metallic ring has the spin degree of freedom.

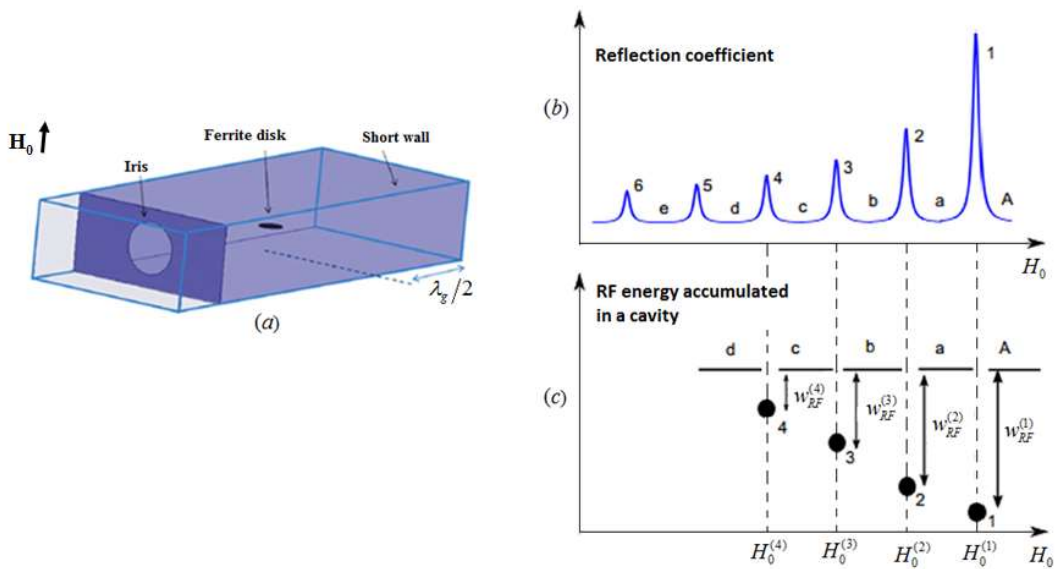

Fig. 13. A sketch showing the relationship between quantized states of microwave energy in a cavity and magnetic energy in a ferrite disk. (*a*) A structure of a rectangular waveguide cavity with a normally magnetized ferrite-disk sample. (*b*) A typical multiresonance spectrum of modulus of the reflection coefficient. (*c*) Microwave energy accumulated in a cavity; $w_{RF}^{(n)}$ are jumps of electromagnetic energy at MDM resonances.

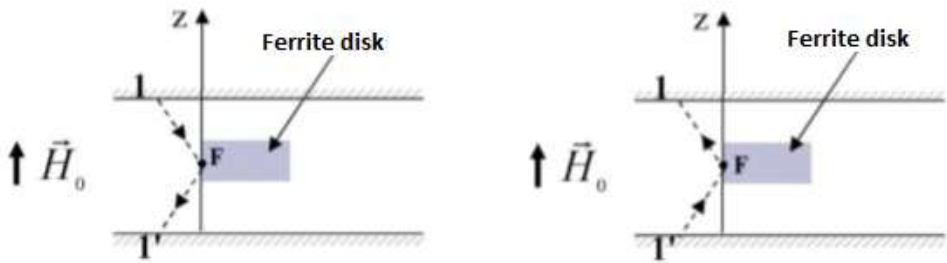

Fig. 14. The rays $1 \to F \to 1'$ and $1' \to F \to 1$ acquire different phases when are reflected by the ferrite.

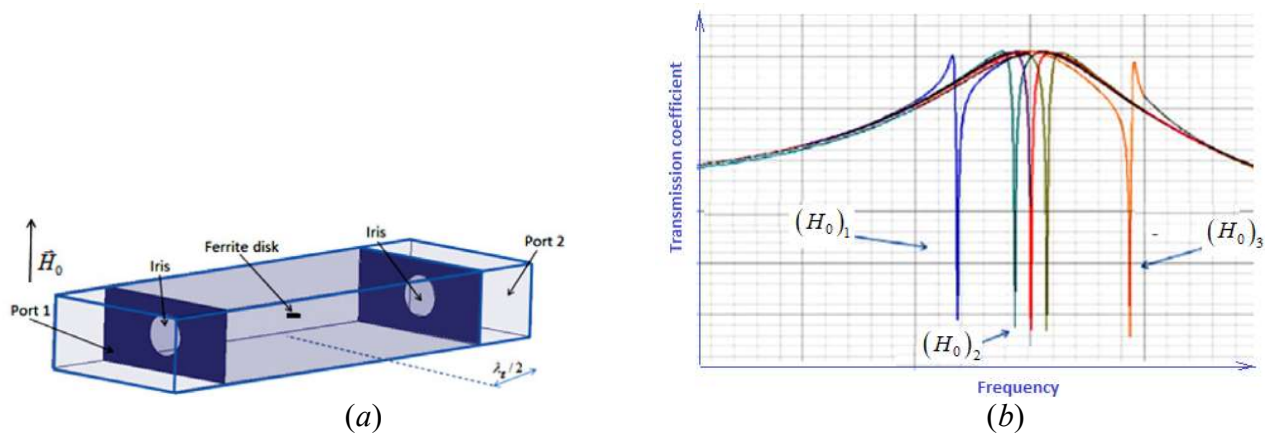

(a)

(b)

Fig. 15. (*a*) A structure of a rectangular waveguide cavity. (*b*) Modification of the Fano-resonance shape. At variation of a bias magnetic field, $(H_0)_1 < (H_0)_2 < (H_0)_3$, the Fano line shape of a MDM resonance can be completely damped. The scattering cross section of a single Lorentzian peak corresponds to a pure dark mode.

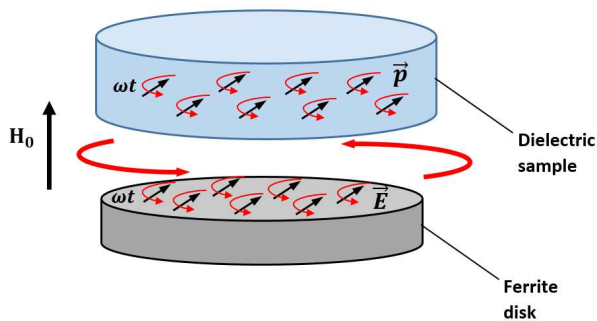

Fig. 16. Angular momentums (spin and orbital) transfer to electric polarization in a dielectric sample.

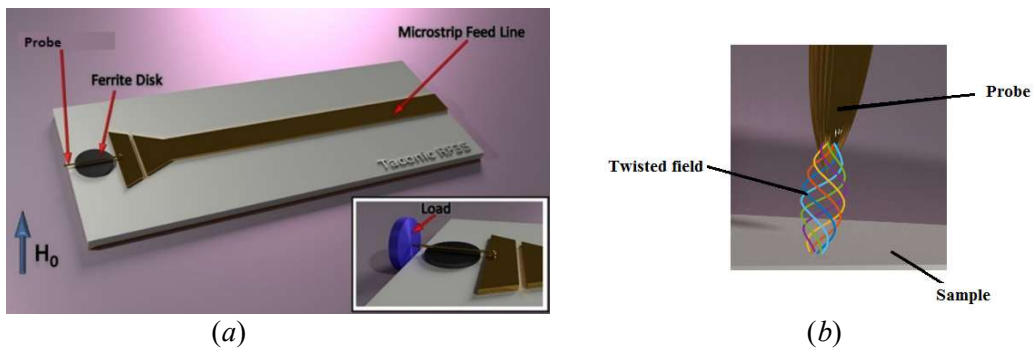

(a)

(b)

Fig. 17. (*a*) Microwave probing structure with a MDM ferrite-disk resonator and a wire electrode. (*b*) Schematic illustration of twisted near fields.

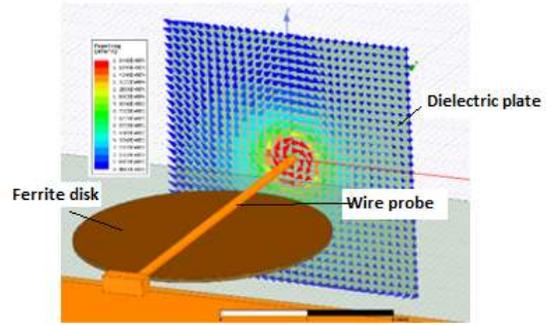

Fig 18. Angular momentum transfer to localized regions in dielectric samples.

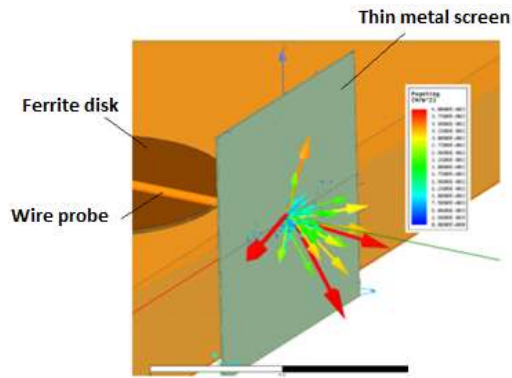

Fig. 19. Numerical results showing the twisted-field effect of penetration microwave power through a thin metal screen.